\documentclass[a4paper,11pt]{article}
\usepackage{pos}
\usepackage{hyperref}

\title{PRECISE localizations of repeating Fast Radio Bursts}

\manuallySeparateAuthors  
\author*[a]{Benito Marcote,}
\author[b]{\ Franz Kirsten,}
\author[c,d]{\ Jason W.~T. Hessels,}
\author[c,d]{\ Kenzie Nimmo,}
\author[a]{\ and Zsolt Paragi}
\author{\ for the PRECISE project}

\affiliation[a]{Joint Institute for VLBI ERIC,\\
	Oude Hoogeveensedijk 4, 7991 PD Dwingeloo, The Netherlands
}
\affiliation[b]{Department of Space, Earth and Environment, Chalmers University of Technology,\\
	Onsala Space Observatory, 439 92, Onsala, Sweden
}
\affiliation[c]{ASTRON, Netherlands Institute for Radio Astronomy,\\
	Oude Hoogeveensedijk 4, 7991 PD Dwingeloo, The Netherlands
}
\affiliation[d]{Anton Pannekoek Institute for Astronomy, University of Amsterdam,\\
	Science Park 904, 1098 XH, Amsterdam, The Netherlands
}

\emailAdd{marcote@jive.eu}

\abstract{
    Fast Radio Bursts (FRBs) are extremely luminous and brief signals (with duration of milliseconds or even shorter) of extragalactic origin. Despite the fact that hundreds of FRBs have been discovered to date, their nature still remains unclear. Precise localizations of FRBs can unveil their host galaxies and local environments -- and thus shed light on the physical processes that led to the burst production. However, this has only been achieved for a few FRBs to date.

    The European VLBI Network (EVN) is currently the only instrument capable of localizing FRBs down to the milliarcsecond level. This level of precision was critical to associate the first localized FRB, 20121102A, to a star-forming region in a low-metallicity dwarf galaxy and physically related it to a compact persistent radio source. Analogously, a second repeating FRB, 20180916B, was found to just outside the edge of a prominent star-forming region of a nearby spiral galaxy.

    The PRECISE project (Pinpointing REpeating ChIme Sources with EVN dishes), starting from 2019, has observed hundreds of hours per year with a subset of EVN telescopes with the goal of localizing repeating FRBs discovered by the CHIME/FRB Collaboration. The ultimate goal of PRECISE is to disentangling the environments where FRBs can be produced.

    Here we present the state of the art of the FRB field, the PRECISE project, and the localizations achieved until now, which have unveiled a variety of environments where FRBs can be found that challenges the current models.
}

\FullConference{
  *** European VLBI Network Mini-Symposium and Users' Meeting (EVN2021) ***\\
  *** 12-14 July, 2021 ***\\
  *** Online ***
}

\begin{document}
\maketitle

\section{Introduction}

The transient sky holds a diversity of astronomical phenomena that occur in extreme environments where the physical conditions suddenly change. These transient events involve a plethora of different environments and physical phenomena, and their timescales can range from fractions of seconds to years.
Frequently, the discovery of new types of transient events occurs when we explore a new region in the observing parameter space, which remained unexplored up to that moment. Technical developments and new observatories, or new approaches implemented in the data analysis, have thus always opened new astrophysical fields.

Fast Radio Bursts (FRBs) are exceptionally luminous transient sources of unknown physical origin that last for less than thousandths of a second \cite{petroff2019}. To date FRBs have only been detected in the radio domain and most of them were found at cosmological distances. FRBs are not rare events, the currently best-estimated rates predict $\sim 10^{3\text{--}4}\ \mathrm{FRBs\ sky^{-1}\ d^{-1}}$ \cite{keane2015}. However, their brief emission combined with being one-off events made them to remain unknown to us until 2007 \cite{lorimer2007}. Among all discovered FRBs, only a fraction of them are (apparently) capable of producing multiple bursts and thus {\em repeat}.

Despite hundreds of them have been discovered to date, only tens of them have been associated to host galaxies. And even a smaller number of them have been localized with enough precision to unveil their local environments. The nature of FRBs still remains unclear, although hundreds of models have been postulated to date (see \url{frbtheorycat.org} or \cite{platts2019}).

FRB studies may have remarkable implications for a wide variety of different research  fields, in particular due to their cosmological distances and the short time scales of the emission. For instance, they can allow astronomers to constrain the baryon content of the Universe \cite{macquart2020}, the Hubble constant \cite{wu2021}, tracing the properties of the intergalactic medium and Galactic Halos \cite{prochaska2019haloes}, or put strong constraints on fundamental physics as the equivalent principle \cite{bertolami2018}.

\section{Pin-pointing FRBs with high angular resolution}

A large number of FRBs has already been discovered. However, we know little about the vast majority of them due to their poor localizations. Aside statistical studies to relate FRBs to other known types of transient events \cite{bhandari2022}, the only way to unveil the true nature of FRBs is to localize them with enough precision so one can study in detail the local environments where the bursts are produced. The properties of these environments could provide important clues about the physical origin of FRBs.

Due to the cosmological distances to most FRBs, very high resolution observations are mandatory to fulfill these attempts. Very long baseline interferometry (VLBI) observations are the only suitable technique to reach such resolution and thus to allow us to pin-point the FRBs within their host galaxies.
To date, the European VLBI Network (EVN) has been the only VLBI instrument capable of pinpointing FRBs on milliarcsecond precision.

Despite the first discovery of a FRB occurred in 2007 \cite{lorimer2007}, it took ten years until a FRB was precisely localized and its host galaxy was identified. FRB~20121102A, the first FRB known to repeat \cite{spitler2014}, was found to be located inside a star-forming region of a low-metallicity dwarf galaxy at a redshift of 0.193 \cite{chatterjee2017,marcote2017,tendulkar2017,bassa2017}. Furthermore, the FRB was spatially consistent with a compact ($< 0.7~\mathrm{pc}$ in size) $\sim 200$-$\mathrm{\upmu Jy}$ persistent radio source of still unclear origin \cite{marcote2017}.

A second repeating FRB, 20180916B, was later localized to a spiral galaxy at a redshift of 0.0337, near a region of star formation \cite{marcote2020}. Later observations revealed the bursts to be produced right at the edge of the aforementioned region \cite{tendulkar2021}, putting strong constraints on the possible age of the source and the possible scenarios.

While these two precisely localized FRBs were found in places with prominent star-forming regions, the physical conditions in both environments revealed significant differences. This thus did not clarify the ultimate origin of the sources producing FRBs.
Additionally, tens of FRBs localized to the arcsecond level by the ASKAP telescopes, both repeating and non-repeating FRBs, showed that these events can be found in different environments and host galaxies \cite{bannister2019,ravi2019,prochaska2019,chittidi2020,heintz2020}.
Among them, it is worth to note that only a couple of FRBs have been associated with a persistent and compact radio source.

It is thus clear that there could be a variety of sources, exhibiting complex but different properties, that could produce FRBs \cite{marcote2020}.

\section{The EVN and the PRECISE project}

As it has been shown, our team has been following up repeating FRBs to localize them to milliarcsecond scales with the EVN and thus identify their host galaxies and, even more importantly, their local environments.
However, FRB localizations are an exhaustive observing-time-consuming task, given that the arrival times of FRBs cannot be predicted in advance. Therefore, there was a limit on the number of observations that we could trigger on these sources due to the limited observing time within the EVN observing sessions.

To tackle this problem, we created the Pinpointing REpeating ChIme Sources with EVN dishes (PRECISE) project to carry out out-of-session EVN observations with a sub-array of the available telescopes. \footnote{\url{https://www.ira.inaf.it/precise/Home.html}.}.

The CHIME telescope is currently the most prolific machine to discover FRBs. Hundreds of them have been discovered during the last years, and most of them localized with a-priori localizations of $\sim 1\text{--}10~\mathrm{arcmin}$. This uncertainty fits the primary beam of the EVN telescopes at gigahertz frequencies, and thus the positions of the FRBs can be recovered in these observations.
Among them, there is a fraction of FRBs observed to repeat. As the CHIME/FRB Collaboration monitors the sky every day, they can provide reliable estimations of periods when these FRBs show an enhanced activity of bursts.

Following up the repeating FRBs reported by the CHIME/FRB Collaboration (large number of potential FRBs) with the EVN (capable of localizing them to milliarcsecond resolution) is an ideal combination to exploit different sides of the FRB science and understand the local environments where these events are produced.

By making use of the smaller dishes of the EVN and green time that they have, plus individual single-dish proposals, we arranged periodic observations with an ad-hoc network of EVN dishes that allowed us to significantly extend the observing time on repeating FRBs. The schedules are produced and distributed to the stations manually, while we submitted a correlation-only proposal to the EVN which allowed us to correlate the data with the SFXC correlator at JIVE (The Netherlands)\cite{keimpema2015} in the case of detecting FRBs. Hundreds of hours of observing time have been granted in this way during the last few years.

After recording, a single-dish single-pulse search is conducted within the data of the larger dishes.
Currently, PRECISE observations include up to the following antennas: Badary, Onsala, Shanghai, Svetloe, Toru\'n, Urumqi, Westerbork, Zelenchukskaya (through director's discretion time approval), Effelsberg, Irbene, Medicina, Noto, and Sardinia (through dedicated single-dish proposals).

\section{New PRECISE localizations of FRBs}

In the following we summarize a couple of the most remarkable localizations that we have obtained during the last year, shown in Figure~\ref{fig:locs}.

\subsection{The absence of compact persistent emission associated with FRB~20201124A}

FRB~20201124A was a repeating FRB discovered by CHIME that was particularly active in the first half of 2021 \cite{chime2021}.
Several observatories followed up this enhanced activity,  detecting new bursts and localizing them to arcsecond scales \cite{day2021,xu2021,law2021,wharton2021burst}.
Intriguing, on these scales the Karl G. Jansky Very Large Array (VLA) and the upgraded Giant Metrewave Radio Telescope (uGMRT) detected a persistent radio counterpart consistent with the position of the bursts \cite{ricci2021,wharton2021persistent}.
The limited precision on the localization of both the FRB and the persistent radio source (comparable to the angular extension of the host galaxy) did not allow to provide further clarification on the putative association between both sources, and their relative positions within the host galaxy.

We followed up FRB~20201124A within the PRECISE project during different epochs and, we detected 13 and 5 bursts in two of them \citep{marcote2021atel}.
By combining the data from all these bursts, we were able to localize FRB~20201124A with a 1-$\sigma$ uncertainty of 2.7 milliarcseconds (mas) \cite{nimmo2022frb20201124a}. This is roughly a factor of 1\,000 times better than the previously reported localizations (see Figure~\ref{fig:locs}, left).
We then revealed that the bursts are produced off-center of the host galaxy, as in other previously localized FRBs.

More intriguing, the deep continuum maps from the PRECISE observations did not reveal any persistent emission in the field above the rms noise level. The only explanation to accommodate these results with the persistent radio emission reported on arcsecond scales is a star-formation origin. In these case, as later demonstrated, one would expect the emission to be resolved on the VLBI scales (given the limited short baselines typical of the EVN, the emission is completely resolved out in our maps).

It still remains unclear if FRB~20201124A lies inside or at the edge of this star-forming region, which future high-resolution optical observations will reveal. However, the fact that it is another FRB found in the vicinity of significant star formation strengthen the scenarios which assume that FRBs are related to very young systems, likely involving magnetars \cite{margalit2018frb121102}.

\subsection{An FRB inside a Globular Cluster}

FRB~20200120E was initially discovered by the CHIME/FRB Collaboration as a relatively nearby FRB, with a relatively very low DM in the direction of M81 \cite{bhardwaj2021}. This would make it as, by far, the closest extragalactic FRB discovered to date.
The a-priori localization raised four possible candidate hosts for the FRB: a HII region within the M81 galaxy, an X-ray and a radio point-like source, and a globular cluster, also associated with M81.

\begin{figure}[t]
	\includegraphics[width=0.5\textwidth]{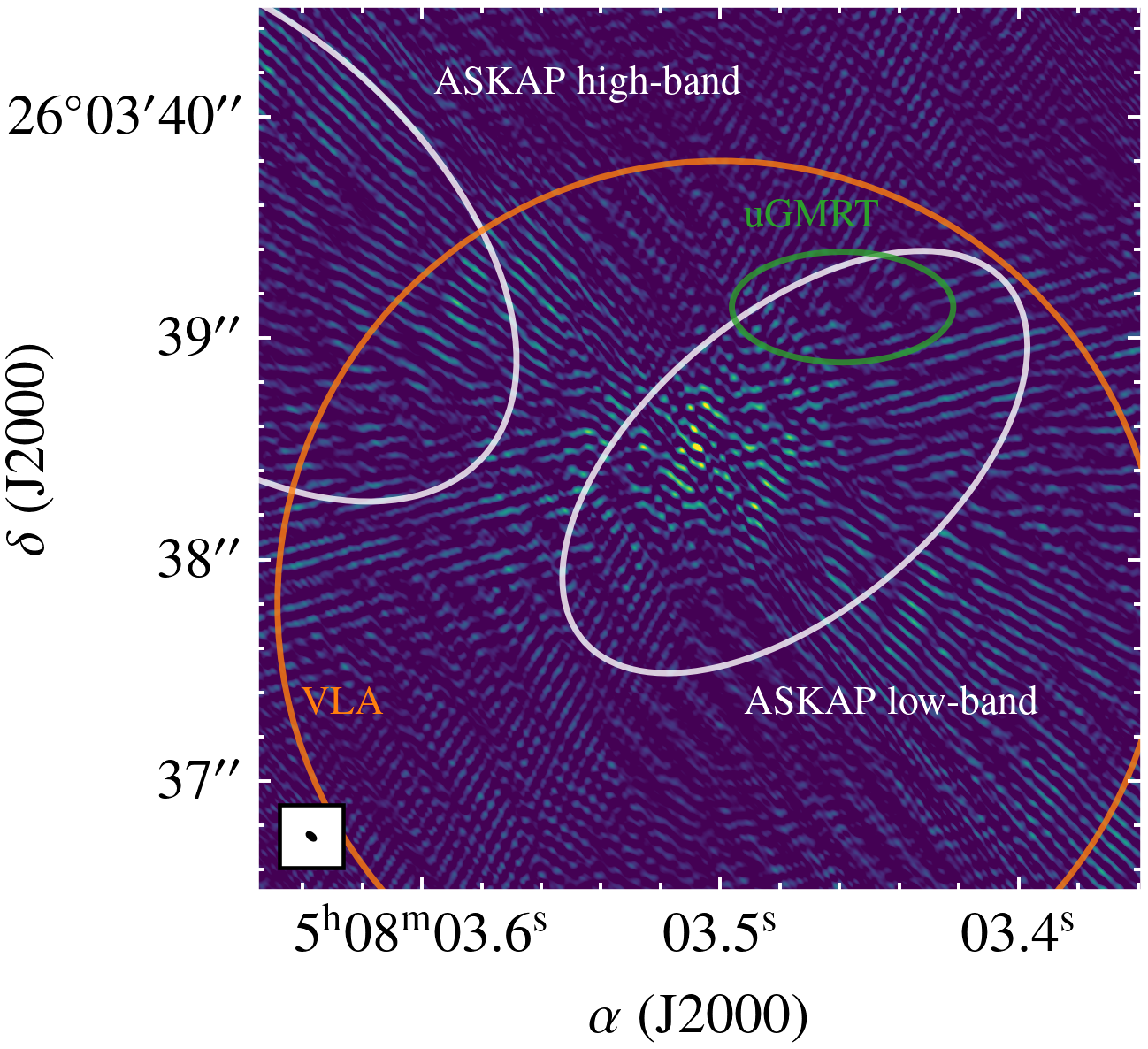}
	\includegraphics[width=0.5\textwidth]{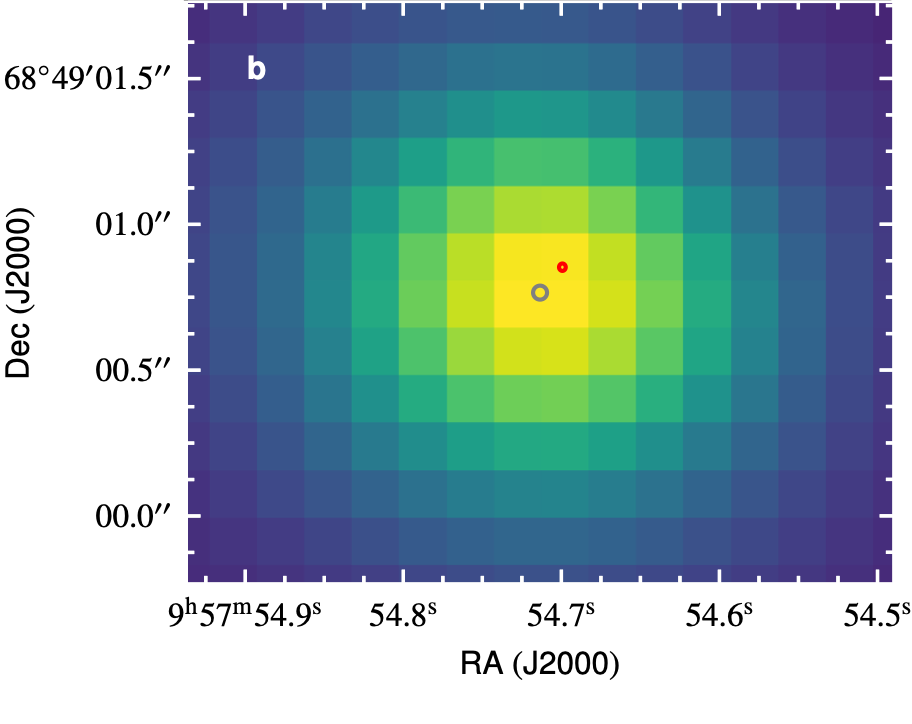}
	\caption{{\em Left}: Dirty image obtained by the EVN by combining the 13 bursts detected from FRB~20201124A. The different ellipses represent the previously reported burst positions. We note that the EVN localization exhibits an improvement of about two orders of magnitude on the localization of FRB~20201124A \cite{marcote2021atel}.
	The synthesized beam is represented by the gray ellipse at the bottom left corner.
	{\em Right:} Optical image of the globular cluster [PR95]~30244 hosting FRB~20200120E. The optical center of light of the globular cluster is represented by the black circle, with its size representing the $3\sigma$ uncertainty. The position of FRB~20200120E is represented by the red ellipse, and its size represents the $10\sigma$ uncertainty. FRB~20200120E is thus significantly offset, by a projected separation of $\approx 2~\mathrm{pc}$.
	}
	\label{fig:locs}
\end{figure}

We detected two bursts from FRB~20200120E during a PRECISE run on 20 February 2021 and another two bursts on 7 March 2021.
These detections allowed us to localize the bursts to a milliarcsecond precision, revealing its origin to inside the globular cluster [PR95]~30244 associated to M81. Furthermore, the achieved precision allowed us to measure its location to $\approx 2~\mathrm{pc}$ offset from the optical center of light of the globular cluster \cite{kirsten2022}, see Figure~\ref{fig:locs}, right.

Given that globular clusters host old stellar populations, all previous models considering FRBs as related to very young magnetars are challenged. However, magnetars can still be formed by other formation channels, such as accretion-induced collapse of white dwarfs or mergers of compact binaries. These scenarios could still reconcile the bursts of FRB~20200120E to be produced a relatively young magnetar in this location.

FRB~20200120E is not unique only due to its unexpected environment. For instance, FRB 20200120E has exhibited the narrowest burst components seen to date, with sub-structures of only $\sim 60~\mathrm{ns}$ of duration \cite{nimmo2022}.
These ultra short timescales imply that the bursts reach brightness temperatures of $\sim 3 \times 10^{41}\ \mathrm{K}$ (without applying relativistic corrections), clearly favoring neutron-star magnetospheric origin.
Additionally, these bursts are comparable to the well-known Crab nano-shots, further strengthening its possible nature as a very young (decades-to-centuries old) system.

\section{Conclusions}

Fast Radio Bursts (FRBs) are a remarkable relatively-new type of astrophysical events of unknown nature, with important implications in different astrophysical, cosmological, or fundamental physics.
Despite hundreds of them have been discovered in the last years, only a few of them have been precisely localized, and associated to their host galaxies and local environments, to date.

Detailed studies of these environments are mandatory to clarify the physical conditions where FRBs can be produced. But only a large number of such precise localizations would allow to statistically analyze these environments. Our group has been localizing FRBs to milliarcsecond precision in the last years thanks to EVN observations. Now, with the PRECISE project, we aim to localize a much larger number of FRBs to fulfill this path.

Ultimately, by unveiling a large number of samples on the local environments of FRBs we would be able to clarify which physical conditions are common to the production of FRBs.

FRB~20201124A seems to be a potential link connecting the Crab pulsar, the Galactic magnetar SGR~1935+2154 (that exhibited FRB-like emission encompassed of X-ray one), and the rest of the FRB population.

The discovery of FRB~20200120E inside a globular cluster in M81 also broke the initial thoughts of having the most active repeating FRBs inside very young environments.

These and previous EVN/PRECISE localizations show the power of having milliarcsecond localizations of FRBs. It is expected that the number of localized FRBs will significantly increase in the coming years and thus we would be able to unveil common properties among these FRBs and conduct population studies.

\bibliographystyle{JHEP}
\bibliography{bibliography.bib}

\end{document}